\documentclass[12pt,preprint]{aastex}
\slugcomment{Draft}

\shorttitle{FIP and Inverse FIP Effects}
\shortauthors{Laming}

\begin{document}

\title{A Unified Picture of the FIP and Inverse FIP Effects}


\author{J. Martin Laming\altaffilmark{1}}


\altaffiltext{1}{E. O Hulburt Center for Space Research, Naval
Research Laboratory, Code 7674L, Washington DC 20375
\email{jlaming@ssd5.nrl.navy.mil}}

\begin{abstract}
We discuss models for coronal abundance anomalies observed in the
coronae of the sun and other late-type stars following a scenario
first introduced by Schwadron, Fisk \& Zurbuchen of the interaction of
waves at loop footpoints with the partially neutral gas. Instead of considering
wave heating of ions in this location, we
explore the effects on the upper chromospheric plasma of the
wave ponderomotive forces.
These can arise as upward propagating waves from the chromosphere
transmit or reflect upon reaching the chromosphere-corona boundary, and
are in large part determined by the properties of the coronal loop above.
Our scenario has the advantage that for realistic
wave energy densities, both positive and negative changes in the
abundance of ionized species compared to neutrals can result, allowing
both FIP and Inverse FIP effects to come out of the model. We
discuss how variations in model parameters can account for essentially
all of the abundance anomalies observed in solar spectra. Expected
variations with stellar spectral type are also qualitatively consistent
with observations of the FIP effect in stellar coronae.
\end{abstract}



\section{Introduction}
Element abundance variations observed in the solar corona and wind with respect
to those determined for the solar photosphere have proved to be one of the
most enduring mysteries of solar physics of the past 15-20 years. The most
commonly observed FIP (First Ionization Potential) Effect refers to the case
where elements with first ionization potential below about 10 eV are
observed to be enhanced in abundance by a factor of about 3-4 in the solar
corona and in the slow speed solar wind. Coronal holes and the fast speed
solar wind which emanates from them by contrast show no such fractionation.

Models for such phenomena have come and gone. Early diffusion models based
on ion coupling to a background flow of protons
\citep{marsch95,peter96,peter98m,peter98}
were shown either to be based on somewhat artificial boundary conditions
\citep{mckenzie97,mckenzie00}, or inherently too slow \citep{henoux95,henoux98}.
The first FIP effect model to include an external forces acting on the plasma ions
that is worked out in some details is that due to \citet{antiochos94}, based on
cross B thermoelectric fields associated with the downward electron flux which
gives rise to chromospheric evaporation. The absence of a FIP effect in coronal
holes arises naturally, but in coronal regions where FIP fractionation occurs,
a mass dependence is predicted which is not observed.
More recently two distinct possibilities have been
discussed in the literature. \citet{arge97} proposed that reconnection
events in the chromosphere heat ions and not neutrals, the higher ion scale
heights then leading to enhanced coronal abundances for those elements
which are ionized in the chromosphere, the low FIP elements. A fractionation
by a factor of 3-4 arises naturally out of this model; if the reconnection
is driven any harder, chromospheric heating ionizes all elements and the
distinction between high and low FIP elements is lost. This might be
problematic, since higher FIP fractionations are observed
in discrete solar features \citep{feldman03}. Less satisfactory is
that the difference between corona/slow wind and coronal holes/fast wind is
not accounted for. In a qualitatively different approach,
\citet{schwadron99} envisaged a wave heated coronal loop
where waves penetrate down to the chromosphere at each footpoint and there
selectively heat the ions but not the neutrals, again leading to a positive
fractionation of the low FIP species. We remark that the
precise nature of the wave heating in their model remains obscure. Ions
are heated to a constant velocity, the wave phase speed, by isotropic
waves. This would suggest a resonant ion cyclotron heating mechanism.
It seems to us that sufficient energy in these waves is unlikely to
reach the partially ionized layer of the chromosphere, due to the
intrinsically small amount of energy generally believed to reside in
high frequency waves, and the damping they would undergo in the partially
ionized chromosphere \citep{depontieu01}. We therefore consider below the
effect of nonresonant waves on chromospheric ions, through the action of
the ponderomotive force that must exist as Alfv\'en waves propagate
through the chromosphere, either as initially upwards propagating waves
impinging on the corona from below, or downward propagating coronal waves
impinging of the chromosphere from above.
The difference between corona/slow
wind and coronal hole/fast wind abundances arises quite naturally in this
model, in that
waves on closed loops inevitably return to loop footpoints unless damped
or reflected higher up, but waves on open field lines do not. In common with
\citet{arge97}, \citet{schwadron99} only produce a positive FIP fractionation,
i.e. only enhancements in abundance of low FIP ions.

This last point has become of great interest with the advent of element
abundance measurements in stellar coronae. \citet{feldman00} reviewed
the observational situation prior to the launch of Chandra and
XMM-Newton. At that time, coronal abundance patterns showed a trend of
increasing metal depletion with increasing activity for the most
active stars, going from FIP effects in single stars of solar-like
activity, through higher activity stars with essentially
no coronal abundance anomaly, right up
to the most active stars with values of the coronal Fe/H abundance
{\em lower} than
solar photospheric values by factors of up to about 3, i.e. a depletion
of an order of magnitude compared to low FIP enhanced coronae like that of
the Sun. The lowest activity solar-like star, Procyon, also showed no
coronal abundance anomaly.
With the advent of data from XMM-Newton and Chandra, the
elemental abundance anomalies in active stellar coronae were revealed
to be more akin to an inverse FIP effect
\citep{brinkman01,drake01a}, in that while high FIP elements
are essentially unchanged relative to H, low FIPs are now depleted by
typical factors of 1/3. The transition from unfractionated to
inverse FIP fractionation with increasing stellar activity is
illustrated by \citet{audard03}. While the ubiquity of the inverse FIP effect
remains controversial in many sources due to uncertainties in the underlying
photospheric abundances (see discussion in section 4), it appears real enough
in at least a few cases to deserve serious consideration. As will be seen in
section 5, an inverse FIP effect can also be produced by ponderomotive forces,
which is an encouraging sign.

\section{The Ponderomotive Force}
The ponderomotive force for Alfv\'en waves has been derived from the term
$\vec{j}\times\vec{B}/c =\left(\nabla\times\vec{B}\right)\times\vec{B}/4\pi$
in the MHD momentum equation by \citet{litwin98}. We work from a more
general expression for
the time-independent ponderomotive force on a particle of
species $s$, $f_j^s$, given by \citep{lee83};
\begin{equation}
F_j^s={1\over 16\pi n_s}\left[\left(K_{\beta\alpha}^s-\delta _{\beta\alpha}\right)
{\partial E_{\alpha}E_{\beta}^{\ast}\over\partial x_j}
+{\partial\over\partial x_l}\left(\epsilon _{jmp}\epsilon _{klm}\Omega _p
{\partial K_{\beta\alpha}^s\over\partial\Omega _k}E_{\alpha}E_{\beta}^{\ast}\right)
-{\partial\over\partial x_l}\left(k_j{\partial K_{\beta\alpha}^s\over\partial
k_l}\right)E_{\alpha}E_{\beta}^{\ast}\right].
\end{equation}
Here $K_{\beta\alpha}^s$ are the terms in the dielectric tensor contributed
by the species $s$, $\vec{E}$ and $\vec{E}^{\ast}$ are the wave electric
field and its complex conjugate, $\epsilon _{jmp}$ is the Levi-Cevita symbol,
$\vec{\Omega}=q\vec{B}/mc$ where $\vec{B}$ is the magnetic field, $q$ and $m$
are the charge and mass of particles of species $s$, $c$ is the speed of light,
and $\vec{k}$ is the wavevector. We will consider the simplest possible
geometry for a loop footpoint for this initial study,
$\vec{B}=\left(0,0,B\right)$ and $\vec{k}=\left(0,0,k\right)$, which means
that $\vec{E}$ has only perpendicular ($x$ and $y$) components for Alfv\'en
or ion cyclotron waves propagating along $\vec{B}$. All plasma quantities
are constant in the region of interest, except for $E_{\perp}$, the
particle density,  and the
ionization fraction of the various elements which may have gradients in the
$z$ direction. This means that $l=p=k$ in the second term in equation 1,
giving $\epsilon _{klm}=0$, and hence this term drops out.

We substitute the form of the dielectric tensor derived using the Vlasov
approach \citep[equation 10.21]{melrose86} to get
\begin{eqnarray}
\nonumber F_j^s&={1\over 16\pi n_s}{4\pi q^2\over m\omega ^2}
\sum _{p=-\infty}^{\infty}\int
{\partial\left|\vec{E}\cdot\vec{V}\right|^2\over\partial z}
{1\over\omega -p\Omega -kv_{\Vert}}
\left({\omega -kv_{\Vert}\over v_{\perp}}{\partial f\over\partial v_{\perp}}
+k{\partial f\over\partial v_{\Vert}}\right)\\
&+{kv_{\Vert}\over\left(\omega -p\Omega -kv_{\Vert}\right)^2}
\left({\omega -p\Omega\over v_{t\Vert}^2}+{p\Omega\over v_{t\perp}^2}\right)
{\partial f\over\partial z}d^3\vec{v}
\end{eqnarray}
where $f=\left(n/v_{t\Vert}v_{\perp}^2\left(2\pi\right)^{3/2}\right)
\exp\left(-v_{\Vert}^2/2v_{t\Vert}^2 -v_{\perp}^2/2v_{t\perp}^2\right)$ is
the species distribution function in terms of parallel and perpendicular
thermal speeds, $v_{t\Vert}=\sqrt{k_{\rm B}T_{\Vert}/m}$ and
$v_{t\perp}=\sqrt{k_{\rm B}T_{\perp}/m}$ respectively,
parallel and perpendicular velocities $v_{\Vert}$, $v_{\perp}$, and
particle density $n_a$. $\left|\vec{E}\cdot\vec{V}\right|^2$ is
evaluated to give $E_x^2v_{\perp}^2\left(J_{p-1}+J_{p+1}\right)^2/4
+E_y^2v_{\perp}^2\left(J_{p-1}-J_{p+1}\right)^2/4$, where
$J_p=J_p\left(k_{\perp}r\right)$ is the Bessel function of order $p$ with
argument $k_{\perp}r\rightarrow 0$ in the MHD approximation.
Evaluating in terms of
$\phi\left(z\right)=-\left(z/\sqrt{\pi
}\right)\int _{-\infty}^{+\infty}\exp\left(-t^2\right)/\left(t-z\right)
dt$, the plasma dispersion function, gives
\begin{eqnarray}
\nonumber F_j^s&={-q^2\over 8m\omega ^2}{\partial E_{\perp}^2\over\partial z}
\left[
{\omega\over\omega -\Omega}\phi\left(\omega -\Omega\over\sqrt{2}kv_{t\Vert}\right)
+{\omega\over\omega +\Omega}\phi\left(\omega +\Omega\over\sqrt{2}kv_{t\Vert}\right)
-\left(1-{v_{t\perp}^2\over v_{t\Vert}^2}\right)\left\{
\phi\left(\omega -\Omega\over\sqrt{2}kv_{t\Vert}\right)+
\phi\left(\omega +\Omega\over\sqrt{2}kv_{t\Vert}\right)-2\right\}\right]\\
\nonumber &
+{q^2E_{\perp}^2\over 8n_sm\omega ^2}{\partial n_s\over\partial z}
\left\{{\omega v_{t\perp}^2\over v_{t\Vert}^2}+\Omega\left(1-{v_{t\perp}^2\over
v_{t\Vert}^2}\right)\right\}\left\{{\omega -\Omega\over k^2v_{t\Vert}^2}\phi
\left(\omega -\Omega\over\sqrt{2}kv_{t\Vert}\right) -
{\omega -\Omega\over k^2v_{t\Vert}^2}-\phi
\left(\omega -\Omega\over\sqrt{2}kv_{t\Vert}\right){1\over\omega -\Omega}\right\}\\
& +{q^2E_{\perp}^2\over 8n_sm\omega ^2}{\partial n_s\over\partial z}
\left\{{\omega v_{t\perp}^2\over v_{t\Vert}^2}-\Omega\left(1-{v_{t\perp}^2\over
v_{t\Vert}^2}\right)\right\}\left\{{\omega +\Omega\over k^2v_{t\Vert}^2}\phi
\left(\omega +\Omega\over\sqrt{2}kv_{t\Vert}\right) -
{\omega +\Omega\over k^2v_{t\Vert}^2}-\phi
\left(\omega +\Omega\over\sqrt{2}kv_{t\Vert}\right){1\over\omega +\Omega}\right\}.
\end{eqnarray}
We always have $\omega +\Omega >>\sqrt{2}kv_{t\Vert}$ so $\phi
\left(\omega +\Omega\over\sqrt{2}kv_{t\Vert}\right)\simeq 1+k^2v_{t\Vert}^2/
\left(\omega +\Omega\right)^2 + 3k^4v_{t\Vert}^4/
\left(\omega +\Omega\right)^4 + ...$. Far from a
resonance we also have $\omega -\Omega >>\sqrt{2}kv_{t\Vert}$ and making the
same approximation for $\phi\left(\omega -\Omega\over\sqrt{2}kv_{t\Vert}\right)$
we get (dropping the subscripts/superscript $s$)
\begin{equation}
F=-{q^2\over 4m\left(\omega ^2 -
\Omega ^2\right)}{\partial\left(E_{\perp}^2\right)\over\partial z}
-{q^2E_{\perp}^2\over 2nm\Omega ^2}{\partial n\over\partial z}
\left({k^2\over\omega ^2}\left(v_{t\Vert}^2-v_{t\perp}^2\right)
+3{k^2\over\Omega ^2}v_{t\Vert}^2\right).
\end{equation}
For Alfv\'en waves, $E_{\perp}^2/8\pi = U_{waves}V_A^2/c^2$ where
$V_A$ is the Alfv\'en speed. Hence $\partial E_{\perp}^2/\partial z
=8\pi U_{waves}/c^2 \partial V_A^2/\partial z =-E_{\perp}^2/n \partial n/
\partial z$, so the three terms in equation 4 are in the ratio
$1$ : $2\left(v_{t\Vert}^2-v_{t\perp}^2\right)/V_A^2$ :
$2v_{t\Vert}^2\omega ^2/V_A^2\Omega ^2$.
For low frequency nonresonant waves the first term
always dominates, and hereafter we neglect the second two.
As $\omega\rightarrow 0$, we recover the expression of \citet{litwin98}.
For wave intensity increasing upwards in the
loop footpoint, the first term gives a force directed {\em downwards}
on the ions if
$\omega >>\Omega$, and {\em upwards} if $\Omega >> \omega$.
The upwards force is proportional to mass, so the resulting acceleration
is mass independent. The reason for such forces lies in the refraction
of waves in a density gradient. All waves are refracted towards plasma
regions with higher refractive index, which means regions of high density
for low frequency waves and regions of low density for high frequency
waves. The increased wave pressure in the region of high refractive
index produces a force which pushes ions in the plasma towards regions
of lower refractive index. This means that low frequency waves push
ions to low density regions, i.e. upwards in a gravitationally
stratified medium, and high frequency waves push ions towards high
density region, i.e. downwards. Close to a resonance  $\phi
\left(\omega -\Omega\over\sqrt{2}kv_{t\Vert}\right)\simeq
\left(\omega -\Omega\right)^2/k^2v_{t\Vert}^2-
\left(\omega -\Omega\right)^4/3k^4v_{t\Vert}^4 + ...$, using
$\omega -\Omega <<\sqrt{2}kv_{t\Vert}$, and so
\begin{equation}
F^s={q^2\over 4m\Omega ^2}\left[{\partial E_{\perp}^2\over\partial z}\left(
{3\over 4}-{v_{t\perp}^2\over 2v_{t\Vert}^2}\right) -{E_{\perp}^2\over n}
{\partial n\over\partial z}{v_{t\perp}^2\over v_{t\Vert}^2}\right].
\end{equation}
Both terms in general give an upwards force on the ions, unless $T_{\Vert}
>> T_{\perp}$. Both positive and negative changes in the abundances of
ions compared to those of neutrals in the partially ionized region of the
chromosphere are possible in the scenario we discuss, but are generally
negligible compared to the ponderomotive force due to nonresonant waves.
In chromospheric plasma, waves of
sufficiently high frequency to produce a downwards ponderomotive force or
to resonate with ion gyrofrquencies
are rapidly damped by charge exchange collisions \citep{depontieu01},
and so their effect on abundance fractionation is negligible.

\section{The FIP Effect}
\subsection{Formalism}
We follow in part the approach and notation of \citet{schwadron99}.
Consider first the motion of ions and neutrals of element $s$
in a background flow of protons and hydrogen
with speed $u$. We neglect the ambipolar force which is generally
much less than gravity, and assume a flux tube of constant cross
sectional area and magnetic field with height
\citep[see e.g.][]{klimchuk00,watko00}
to write the momentum equations for ions and neutrals as
\begin{eqnarray}
{\partial P_{si}\over\partial z}&=-\rho _{si}g-\rho _{si}\nu _{si}
\left(u_{si}-u\right) \\
{\partial P_{sn}\over\partial z}&=-\rho _{sn}g-\rho _{sn}\nu _{sn}
\left(u_{sn}-u\right),
\end{eqnarray}
where $P_{si}$ and $P_{sn}$ are the partial pressures of ions and neutrals
of element $s$, $\rho _{si}$ and $\rho _{sn}$ are the corresponding densities,
$\nu _{si}$ and $\nu _{sn}$ the collision rates with ambient gas (assumed
hydrogen and protons), $u_{si}$ and $u_{sn}$ the flow speeds, and $u$ the
hydrogen flow speed imposed on the loop. We also neglect an inertial
term $\partial /\partial z\left(\rho _su_s^2/2\right)$ since the flow
speed is much lower than particle thermal speeds. The momentum equations
can be combined to give
\begin{equation}
{\partial P_s\over\partial z}=-\rho _sg
-\nu _{eff}\rho _s\left(u_s-u\right)+{\partial\xi _s\over\partial z}
{\rho _sv_s^2\over 2}{\nu _{si}-\nu _{sn}\over \left(1-\xi _s\right)
\nu _{si} +\xi _s\nu _{sn}},
\end{equation}
with $\nu _{eff}=\nu _{si}\nu _{sn}/\left(\xi _s\nu _{sn} +\left(1-\xi _s\right)
\nu _{si}\right)$  and $\xi _s$ the ionization fraction of element $s$.
For
\begin{equation}
u_s=u-g/\nu _{eff}\left(
1-\mu /m_s\right) +\partial\xi _s/\partial z
\left(\rho _sv_s^2/2\right)\left(\nu _{si}-\nu _{sn}\right)/\left\{
\left(1-\xi _s\right)
\nu _{si} +\xi _s\nu _{sn}\right\}
\end{equation}
where $\mu$ is the mean molecular weight, all elements are lifted by the
background flow to the same scale height given by $k_{\rm B}T/\mu g$.
This obviously requires $u\nu _{eff} > g$ (assuming the term in
$\partial\xi _s/\partial z$ negligible).
For $u\nu _s << g$, we get gravitationally stratified solutions with
$\rho _s\propto\exp\left(-m_sgz/k_{\rm B}T\right)$.
With $g=2.74\times 10^4$ cm s$^{-2}$, and $\nu _s\sim 10^2 - 10^3$ the
former case is valid for the Sun for flow speeds in the chromosphere
greater than a relatively modest 10 - 100 cm s$^{-1}$. The solar wind
particle flux at $1 R_{\sun}$ is of order $10^{13}$ cm$^{-2}$s$^{-1}$,
which probably requires a flow speed well in excess of $10^3$ cm s$^{-1}$ at
a density of $10^{10}$ cm$^{-3}$ (accounting for an unknown area
filling factor) to supply it, so we do not expect gravitational separation
of elements in the chromosphere.

The solar chromosphere is doubtless a more dynamic environment than
represented by equations 6-9. For our purposes the net result
of this dynamic behavior is merely to completely mix up the plasma to give
uniform elemental composition with height, which is obtained in our
model with the above choice for $u_s$. Other choices may be possible
which would provide chemical fractionation in the unperturbed
chromosphere, and one could choose $u_s$ to provide the required FIP
effect. However the physics behind such a specification for $u_s$ in
most cases remains obscure, and is probably unrealistic, leading to
an unsatisfactory explanation for the FIP effect. Problems of this
sort abound in models where no external force provides the FIP
fractionation, as in e.g. \citet{marsch95}. Our model starts with
a fully mixed chromosphere, upon which pondermotive forces due to
Alfv\'en wave reflection and transmission
act to provide the fractionation. The low solar chromosphere is of
much higher density than the upper layers where the FIP fractionation
will occur in our models. Consequently the lower boundary condition
of completely mixed photospheric composition material gives an
essentially infinite particle ``reservoir'' to supply the extra
fractionated elements.

We now include a ponderomotive force, $\rho _{si}a +
b{\partial\rho _{si}/\partial z}$ (see equations 4 and 5), on the ions
in the momentum equations;
\begin{eqnarray}
{\partial P_{si}\over\partial z}&=-\rho _{si}g-\rho _{si}\nu _{si}
\left(u_{si}-u\right) +\rho _{si}a +b{\partial\rho _{si}\over\partial z}\\
{\partial P_{sn}\over\partial z}&=-\rho _{sn}g-\rho _{sn}\nu _{sn}
\left(u_{sn}-u\right).
\end{eqnarray}
Taking $u_s$ as specified above and assuming
$\partial\left(\xi _sb\nu _{eff}
/\nu _{si}\right)/\partial z\simeq 0$ we find
\begin{equation}
{\rho _s\left(z_u\right)\over\rho _s\left(z_l\right)}=
{v_s^2\left(z_l\right)+2b\xi _s\nu _{eff}/\nu _{si}\over
v_s^2\left(z_u\right)+2b\xi _s\nu _{eff}/\nu _{si}}
\exp\left\{2\int _{z_l}^{z_u}
\xi _sa\nu _{eff}/\nu _{si}/\left(v_s^2+2b\xi _s\nu _{eff}/\nu _{si}
\right)dz\right\}.
\end{equation}
A quantitative assessment of coronal
element abundances anomalies requires an evaluation of equation 12 with a
realistic model chromosphere in the region of Alfv\'en wave reflection.
With $b=0$, the fractionation produced by the ponderomotive force
is proportional to $\exp\left\{2\int _{z_l}^{z_u}
\xi _sa\nu _{eff}/\nu _{si}/v_s^2dz\right\}$ and is approximately
mass independent if $v_s$ is dominated by the microturbulent velocity,
as it must be if the unperturbed chromosphere is completely mixed with
gravitational scale height corresponding to the mean molecular mass.
Henceforward we take $b=0$, following the discussion in section 2.

\subsection{Alfv\'en Wave Reflection}
The chromospheric reflection and damping of Alfv\'en waves has recently
been studied in some detail \citep{ofman02,depontieu01} in connection with
observations by TRACE of the damping of loop oscillations \citep{nakariakov99}.
Waves with frequency below the ion-neutral charge exchange rate propagate
essentially undamped through chromospheric plasma, but can be reflected at
the chromosphere-corona boundary.

When the Alfv\'en wavelength is much larger than the characteristic length
scale over which the density changes, the amplitude reflection coefficient
is given approximately by $B_r/B_i=\left(V_{A1}-V_{A2}\right)/
\left(V_{A1}+V_{A2}\right)$, where $V_{A1}$ and $V_{A2}$ are the Alfv\'en speeds
of the incident and transmitted wave respectively.
Since $B$ is continuous across the boundary,
$B_t=B_r+B_i$ and $B_t/B_i=2V_{A1}/\left(V_{A1}+V_{A2}\right)$. These
simple results are recovered in the long wavelength limit when the
chromosphere is treated as an exponential atmosphere
\citep{leroy80,ofman02}. The wave magnetic field in this case
can be expressed in terms of Hankel fuctions
\citep{hollweg84,depontieu01} with argument $\left(2h\omega/ V_{Ac}\right)
\exp\left(-z/2h\right)$, where $h$ is the chromosphere scale height,
$V_A=V_{Ac}\exp\left(z/2h\right)$ is the chromospheric Alfv\'en speed, in
terms of its (assumed constant) coronal value $V_{Ac}$ and
$z \left(<0\right)$ is the
depth below the corona. When $\left(2h\omega/ V_A\right)
\exp\left(-z/2h\right)<<1$ (the long wavelength limit)
the expression can be simplified by expressing the Hankel functions
in terms of Bessel functions and taking the appropriate limits when the
arguments are much less than unity. The result is
\begin{equation}
B_{\perp}={irB_0\over V_{Ac}}\left[\left(e+f\right){\alpha\over 2}
\exp\left(-z/h\right) -i\left(e-f\right){2\over\pi\alpha }
\right]\exp\left(i\omega t\right),
\end{equation}
where $\alpha = 2h\omega/V_{Ac}$,
$B_0$ is the longitudinal (static) magnetic field, $r$ is the
radius of the assumed torsional oscillation, and
$e$ and $f$ are complex constants
representing the magnetic field amplitude of the upward and downward
propagating chromospheric waves respectively.
So long as the Alfv\'en wave energy flux through the chromosphere is
nonzero (i.e. $e\ne f$), with $\alpha\sim 10^{-3}-10^{-2}$ for
solar parameters, $B_{\perp}$ and the wave energy density
$B_{\perp}^2/8\pi$ are essentially independent of $z$ in all regions
except the lowest chromospheric levels.
Making the identification $U_{waves}=B_{\perp}^2/8\pi
=c^2E_{\perp}^2/8\pi V_A^2$, for the wave energy density we deduce
\begin{equation}
{\partial E_{\perp}^2\over\partial z}={V_A\over c^2}16\pi U_{waves}
{\partial V_A\over\partial z}=
{V_A^2\over c^2}{8\pi U_{waves}\over h}.
\end{equation}
The ponderomotive acceleration
is then $a=U_{waves}/2h\rho\simeq 2.5\times 10^4$ cm s$^{-2}$ for
$U_{waves}=0.1$ erg cm$^{-3}$, $h=200$ km and $\rho = 10^{-13}$
g cm$^{-3}$. More realistic chromospheric models employed below
\citep{vernazza81} have steeper density gradients in certain
regions, giving even stronger ponderomotive acceleration. If sufficient
cancelation exists between $e$ and $f$ so that the first term in
equation 13 dominates, $B_{\perp}\propto\exp\left(-z/h\right)$ and
$E_{\perp}\propto\exp\left(-z/2h\right)$, giving a ponderomotive
force on ions directed downwards. The possibility that solutions of
this type might play a role in the inverse FIP effect is discussed in
more detail below.

\subsection{Simulations}
\begin{figure}
\epsscale{0.75}
\plotone{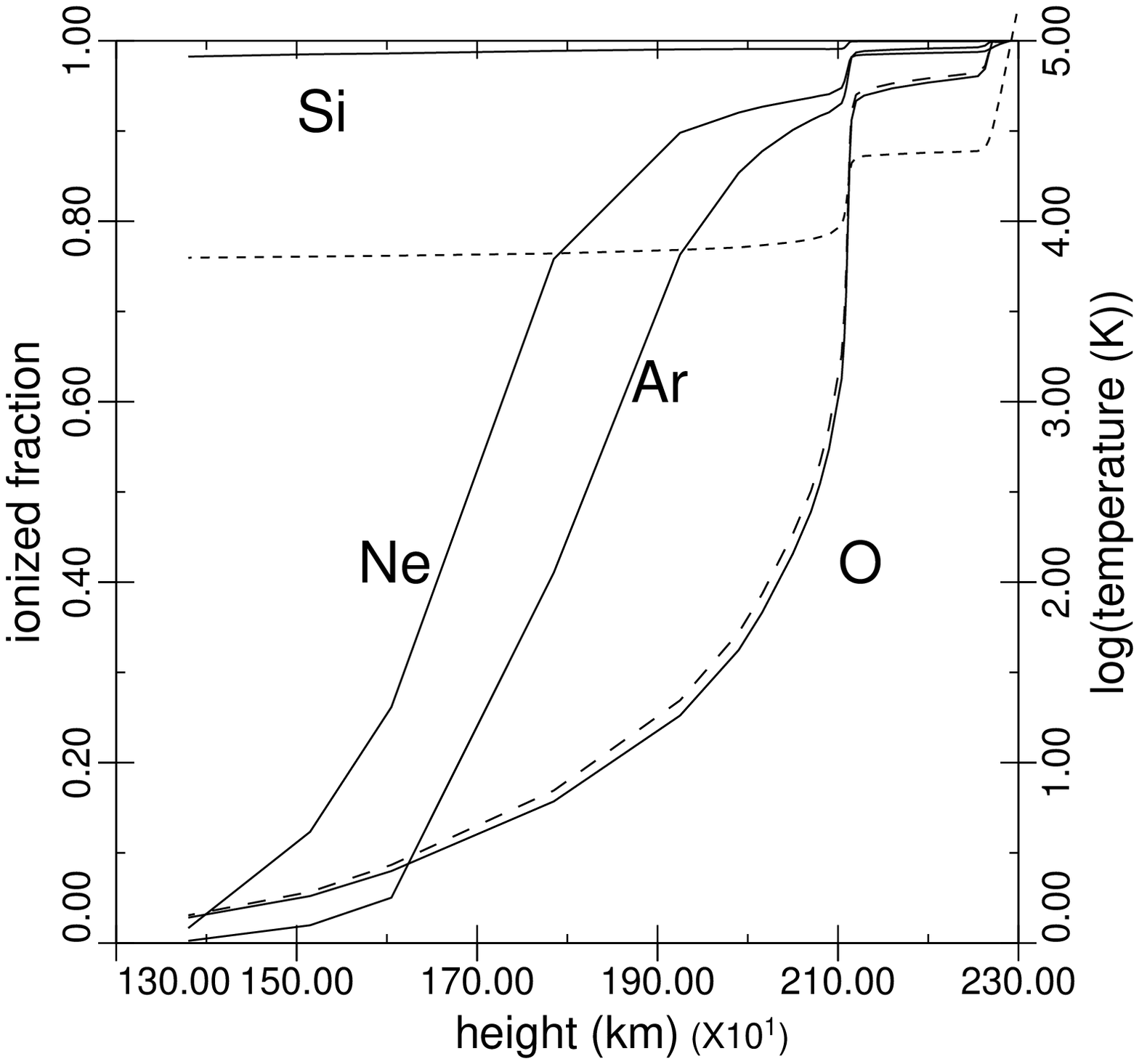}
\figcaption[f1.eps]{Plots of ionization fraction (solid lines) of
Si, O, H, and Ne against height above the photosphere
derived from model calculations. The long dashed line shows the H ionized
fraction
in the VALC model for comparison. The short dashed line, to be read on
the right hand
y-axis, gives the plasma temperature from VALC.\label{fig1}}
\end{figure}

We calculate the wave energy gradient using the results above in the model
chromospheres of \citet{vernazza81}. Atomic data for ionization and recombination
rates are taken from the compilation of \citet{mazzotta98}, including rates
due to charge transfer recombination and ionization from
\citet{kingdon96}. Photoionization rates are evaluated
using fits to cross sections in \citet{verner96}. At each layer of the
chromosphere, the average solar spectra of \citet{vernazza78} and
\citet{malinovsky73} are attenuated
by absorption by atomic hydrogen, and then used to compute the photoionization
rates for neutral species, which together with collisional ionization and
recombination rates are used to calculate the ionization fractions of the
various elements. Plots of the ionization fraction of O, Ne, Si, and Ar against
height above the photosphere coming from our models are given as
solid lines in Figure \ref{fig1}. The long dashed line gives the H ionization
fraction in the VALC model, and the short dashed line gives the temperature
in the model. All other low FIP elements are virtually indistinguishable
on this plot to Si, i.e. all retain ionized fractions very close to unity
for the range of heights shown. The O ionization balance follows that of
H very closely because of the fast charge exchange rates between O and H.
This is due to the close correspondence between their first ionization
potentials.

We evaluate equation 12 integrating the VALC model, corresponding to
average quiet sun, through the chromosphere. The ponderomotive
acceleration is calculated from the gradient of the Alfv\'en speed
in the VALC chromospheric model and equation 14. This is not completely
self-consistent, since equation 14 is derived assumed an exponential
atmosphere, not the VALC model, but should be adequate for the purposes
of this paper.
The ion-proton elastic collision rate is given by the standard Spitzer
formula and is numerically:
\begin{equation}
\nu _{\rm ion-p}={3.1\times 10^4\over A}\left(T\over 10^4 {\rm K}\right)^{-3/2}
\left(n_p\over 10^{10} {\rm cm}^{-3}\right) \quad{\rm s}^{-1}.
\end{equation}
Collision rates where one or both particles are neutral are calculated from the
effective cross sections given in
\citet[and tabulated for reference in Table 1]{vauclair85}, assumed constant with
proton or hydrogen velocity.
Formally, such behavior arises in the limit that the scattering particle
wavefunction is much larger than the range of the scattering potential
\citep[see e.g.][]{landau77}. Protons
at $10^4$ K have de Broglie wavelengths of about 4 \AA\ , to be compared with
typical atomic size or potential ranges of about 2 \AA\ . The numerical value
is
\begin{equation}
\nu _{\rm ion-H}\simeq
{9.1\sigma _{15}\over A}\left(T\over 10^4 {\rm K}\right)^{1/2}
\left(n_H\over 10^{10} {\rm cm}^{-3}\right) \quad{\rm s}^{-1},
\end{equation}
where $\sigma _{15}\sim 1$ is the scattering cross section in units of
$10^{-15}$ cm$^2$. Previous authors
\citep{vonsteiger89,marsch95,schwadron99} have used formulae
for ion-hydrogen and proton-neutral collision rates involving the static atomic
polarizability which give collision rates larger by a factor typically about 2.
In equations 6-11, $\nu _{si}=\nu _{ion-p}+\nu _{ion-H}$ and
$\nu _{sn}=\nu _{ion-H}\left(1+n_p/n_H\right)$. We estimate $\nu _{sn}$ for
other elements by scaling from values given in Table 1 using tabulated values
of static polarizabilities for neutral atoms of the various elements in
\citet{lide95}.

We give in Table 2 the FIP fractionation for a variety of elements
computed for
the VALC background model for Alfv\'en wave energy densities ranging between
0.01 and 0.1 ergs cm$^{-3}$, corresponding to non-thermal mass motions of
up to 10 km s$^{-1}$ at a density of $10^{11}$ cm$^{-3}$. At an energy
density of 0.04 ergs cm$^{-3}$, we find a very
encouraging correspondence between our model values
and observational values taken from the reviews of
\citet{feldman00} and \citet{feldman03}. In particular an almost mass
independent fraction of around 3 occurs for Mg, Si, Fe, and Ni.
The very low FIP and extremely reactive elements Na and K show
fractionations of 5.42 and 7.22 respectively, also
consistent with observations. Of the high FIP elements
(excluding S), Ar has the highest abundance enhancement of 1.25. S is
something of a special case, being the high FIP element with the lowest
FIP of 10.36 eV. Our model gives it an intermediate behavior, with a
FIP fractionation of 1.93. Al and Ca are predicted to have similar mass
independent fractionation to Mg, Si, and Fe, whereas observations indicate
a behavior more like Na and K. This discrepancy would probably be resolved
by improved atomic data for neutral-H collisions for these elements.
We emphasize that given the background
model and the atomic data, there is {\em only one free parameter}
in all these models, namely the Alfv\'en wave energy density.
Insignificant fractionation between H and He occurs in our model. The abundance
ratio He/H has recently been measured in regions of quiet solar corona
\citep{laming03} to be similar to that observed in the slow speed solar
wind, i.e. in the range 0.04-0.05 instead of the value inferred for the
solar envelope by helioseismology of 0.085 \citep{basu98,kosovichev97}. The
He/H abundance ratio in the solar wind is observed to be quite variable over
the course of the solar cycle \citep{aellig01a,aellig01b}, and these
variations are not present in the FIP fractionation of other elements.
Consequently, one should not expect the physical mechanisms that produce
the FIP effect to be responsible for the He fractionation.

Different VAL models produce different FIP fractionations. For 0.04
ergs cm$^{-3}$ wave energy density, runs for the Si FIP fractionation
which with VALC is given in Table 2 as 3.42, VALA gives 30.2, VALB
6.59, VALD 2.56, VALE 2.02 and VALF give 1.69. These models range
from a dark point within a cell (VALA), an average cell center (VALB),
average quiet sun (VALC),average network (VALD), a bright network element
(VALE) and a very bright network element (VALF).

We have extended the elements we consider beyond the list of those usually of
interest to spectroscopists to include Kr, Rb and W. Modest increases in the
wave energy density from that which gives the observed quiet sun FIP
fractionation produce enormous increases in the abundances of Rb and W,
which may be of relevance to element abundances in some impulsive solar
energetic particle events. Kr on the other hand, behaves like the other high
FIP elements. The $^3$He abundance enhancements of factors
$10^3 - 10^4$ in such events are relatively well known.
More modest abundance enhancements of heavy elements are also
seen, e.g. Fe is observed enhanced relative to O by a factor $\sim
10$ over usual coronal abundances. This is usually interpreted as
being due to more efficient stochastic acceleration of ions with
lower charge to mass ratios, i.e. those ions with longer
gyroradii. Such acceleration occurs in the flare loop rather than
in the chromosphere.
Going to elements heavier than Fe, even stronger anomalies have
recently been found \citep{reames00}. Measurements with the
Wind/EPACT/LEMT (Energetic Particle Acceleration, Composition and
Transport/Low Energy Matrix Telescope) instrument reveal
elements with $34\le Z\le 40$ overabundant by a factor $\sim 100$
and those with $50\le Z\le 56$ overabundant by $\sim 1000$,
relative to coronal values. These anomalies are similar in magnitude
to those for $^3$He. The ACE/ULEIS (Advanced
Composition Explorer/Ultra Low Energy Isotope Spectrometer)
instrument reveals similar anomalies going out as far as Bismuth
($Z=83$). If these abundance anomalies were related to the FIP effect
under discussion here, we would expect $34 \le Z \le 40$ bin to be
dominated by the very low FIP elements Rb, Sr, Y, etc, with Kr remaining
essentially unchanged and the bin
at $50 \le Z \le 56$ to be dominated by Cs and Ba. Some FIP selectivity
is evident around atomic masses corresponding to Kr, Rb, Sr, etc
in recent observations of SEP ultra-heavy ions in impulsive solar flares
\citep[][their figure 6]{mason04}, but the data appear to be in better
agreement with our computed values for
wave energy densities around 0.04 ergs cm$^{-3}$, rather than the higher
values. Data around Xe, Cs, and Ba are noisier, but show essentially no FIP
selectivity. We emphasize that the degree of fractionation
depends on the assumption that in the chromosphere unperturbed by Alfv\'en
wave reflection, these elements are sufficiently coupled to the
background flow to rise
up to the same gravitational scale height as H and the other elements, and
that this might be questionable for these, the heaviest of the elements
in our sample. Some gravitational settling of heavy elements does appear to
be in evidence in the solar photosphere and convection zone, as indicated
by the disagreement between recent spectroscopic photospheric abundance
measurements and determinations from helioseismology
\citep{bahcall04,asplund04}.

\section{Alfv\'en Wave Reflection Revisited and Variation in the FIP Effect}
Before proceeding to discuss how variations of the FIP fractionation,
including the possibility of an inverse FIP effect, might arise, we consider
in more detail the properties of Alfv\'en waves in the chromosphere.
\citet{hollweg84} modeled the transmission of Alfv\'en waves from
the chromosphere into the corona using a three layer loop model; a
coronal midsection anchored in the chromosphere at each end.
Waves are fed in at one end, may undergo reflection or transmission
each time they encounter a chromosphere-corona boundary, and can leak out
of the other footpoint.
In Figure 3 we show the transmission coefficient into the corona for
upward propagating waves in the chromosphere (lower panel). The
results are calculated for a model with chromospheric scale height
$h=2\times 10^7$ cm, coronal Alfv\'en speed $V_{Ac}=10^8$
cm s$^{-1}$, and loop length $d=10^{10}$ cm (solid curves), and for
a coronal hole where the limit $d\rightarrow\infty $ is taken
(dashed curves; Hollweg's two-layer model). The upper
panel shows the associated wave magnetic field at the corona-chromosphere
boundary, in units of the initial upward wave amplitude $e$ expressed
as a fraction of the unperturbed magnetic field strength $B$ and the
radius of the torsional oscillation $r$. The strongest wave magnetic
field, and hence energy density and ponderomotive force, are clearly
associated with the maxima in the Alfv\'en wave transmission. From the
difference in the plots for a closed loop and the coronal hole, we see
that most of the waves responsible for the FIP fractionation must be
of relatively low frequency close to the loop fundamental near a period
of 200 s, in order for different fractionations to appear in each case.
At higher frequencies corresponding higher order harmonics of the loop,
the difference in wave magnetic field between the closed loop and
coronal hole is rather small, and less obviously capable of producing
different fractionations. The Alfv\'en wave energy
density represented by the plots in Figure 3 is nearly 6 times higher
in the closed loop than in the coronal hole assuming a flat wave spectrum
between 10 and 1000 s periods. In the range between 100 and 1000 s
period, the closed loop has a factor 20 more wave energy density. With
reference to Table 2, these differences are more than sufficient to
produce the observed abundance differences. Note that in each case we
have assumed the same wave amplitude incident from below.

If it is the case that the solar FIP effect in coronal loops is due
to the approximate equality of wave frequencies generated within the
convection zone and hence the chromospheric Alfv\'en wave spectrum
\citep[see e.g.][]{hathaway00,mcateer04}, and the resonant
frequencies of coronal loops, then a number of other predictions should
follow. First, the FIP enhancement in any particular solar loop should
depend on its size. This does indeed appear to be the case, in that the
small loops reaching maximum temperatures below $10^6$ K
\citep[the unresolved fine structures][]{feldman83,feldman87} do have
element compositions resembling the solar photosphere \citep{laming95},
whereas larger higher temperature loops show the usual FIP effect. This
is presumably because these smaller loops have resonant frequencies
too high for efficient Alfv\'en wave transmission to their higher
temperature regions. Further, the FIP effect in stars of different
spectral type might also show interesting variations from the solar case.
Stars of earlier type than the Sun (G2V) will have shallower convection
zones, and are generally thought to have lower coronal magnetic fields.
Consequently the chromospheric Alfv\'en wave spectrum might be expected
to be of higher frequency than that in the Sun, while the resonant
frequencies of coronal loops are lower, more similar to solar coronal holes.
The observed absence of the FIP effect in the corona of Procyon
\citep[F4IV;][]{drake95a,drake95b,raassen02,sanz04}
is indicative of such an example. Going
to later spectral type than the Sun, the deeper convection zones
generate lower frequency chromospheric waves, but also higher coronal magnetic
fields leading to a mismatch in the opposite sense,
i.e. the coronal loop resonant frequency is higher than
the chromospheric wave spectrum leading again to inefficient wave
transmission and reduced abundance enhancements. FIP effects reduced
in magnitude compared to that in the Sun, but not absent, are seen
in $\epsilon$ Eridani \citep[K2V;][]{laming96,sanz04}, and $\alpha$ Cen AB
\citep[G2V and K1V;][]{drake97,raassen03a}, while
$\xi$ Boo A \citep[G8;][]{laming99,drake01b}, $\pi ^1$UMa (G1V) and
$\chi ^1$Ori
\citep[G0V;][]{gudel02} show similar FIP fractionation to the solar corona.

It is less clear what the correspondence between upward propagating
chromospheric wave frequencies and resonant frequencies of coronal
loops should be in interacting binary stars such as RSCVn and Algol-type
binaries. Observationally, the pre-Chandra/XMM-Newton view of
essentially zero coronal abundance fractionation at the low end of the
activity scale for these objects going over to inverse FIP effect
or metal depletion at the high reviewed by \citet{feldman00} still
holds true, with the caveat raised recently by \citet{sanz04},
that fewer and fewer of these stars
seems to exhibit true metal depletion in their coronae, the coronal
abundance merely reflecting metal poor photospheres. However metal
depleted coronae, or inverse FIP effect, does appear to exist in
some cases, e.g.
II Peg \citep[K2IV plus an unseen companion;][]{huenemoerder01},
AR Lac \citep[G and K subgiants in a 1.98 day orbit;][]{huenemoerder03} and
AB Dor \citep[K2 IV-V with a 0.515 day spin period;][]{sanz03}.
Further, the variation of element
abundance during stellar flares, in which initially metal depleted
plasma evolves towards the standard composition, interpreted in
terms of the chromospheric evaporation of unfractionated plasma to
the coronal flare site, seems to require the existence of such abundance
anomalies. Such phenomena are observed
in HR 1099 \citep[K1 IV and G5 IV;][]{audard01},
Algol \citep[B8 V and K2 IV;][]{favata99}, and
UX Ari \citep[G5 V and K0 IV;][]{gudel99} where the later type subgiant is
taken to be the main source of coronal emission, as well as
AB Dor \citep{gudel01a},
YY Gem \citep[dMe and dMe with 0.814 day orbit;][]{gudel01b},
II Peg\citep{mewe97}, and AT Mic \citep[dM4.5 and dM4.5;][]{raassen03b}.

\citet{hollweg84} argues further that some coronal dissipation of the
Alfv\'en waves must exist, in order to overestimating the nonthermal
line broadening. The inclusion of wave damping does not qualitatively
change our considerations, except that for increased damping the
difference in wave properties between coronal holes and closed loops
becomes smaller. Only at wave damping rates beyond the validity of
Hollweg's analytic treatment does this difference actually disappear.

In Figure 4 we plot the variation of the wave electric field (upper panel)
and magnetic field (lower panel) with distance through the chromosphere
below the corona. A positive gradient of wave electric field gives an
upwards ponderomotive force, and is realized in all cases close to the
coronal boundary. At a wave period of 207 s, corresponding to the loop fundamental
frequency,  the upwards force is strongest and is present
at all heights. For most chromospheric heights the off resonant waves,
corresponding to the magnetic field minimum at wave period 408 s
gives decreasing wave electric field with increasing height, which would
lead to a downwards ponderomotive force on the ions. This is discussed
further below in the subsection on the inverse FIP effect. Away from these
two limiting cases, the chromospheric wave fields generally show
negative gradient of electric field at low heights and a positive gradient
higher up. In our simulations above with hopefully more realistic
chromospheric density profiles coming from VAL models, most FIP
fractionation occurred within 200 km of the chromosphere-corona boundary,
and so we should expect a closed loop excited by a broad spectrum of
Alfv\'en waves below to generally show a positive FIP enhancement. However it is
not absolutely clear that the top of the chromosphere in VAL models can be
identified with the chromosphere-corona boundary in Hollweg's model, which
really marks a transition from an exponentially stratified atmosphere to a
uniform density corona, and might reasonably be taken to indicate the top of the
transition region instead.

\section{The Inverse FIP Effect?}
We now consider possibilities arising with wave reflection in the
chromosphere to produce a so-called Inverse FIP effect. We mentioned above
that high frequency waves from the corona capable of producing a
downwards ponderomotive force on ions are likely to be heavily damped
by charge exchange in the partially neutral chromosphere. Even the lower
charge exchange rates in molecular as opposed to atomic gas do not
allow sufficient ponderomotive forces, even for wave energy densities
in the range up to 1 erg cm$^{-3}$. Further, a ponderomotive
acceleration produced by such waves would not in general be mass
independent, in conflict with observations
\citep[see e.g.][]{audard03,sanz04}.

We instead turn our attention to the most direct ``inverse'' of the
model outlined above. Alfv\'en waves coming up from the stellar convection
zone can be reflected back down again unless they are of the precise
frequency to be transmitted all the way into the corona
\citep{hollweg84,depontieu01}. Taking the corona-chromosphere boundary in
Hollweg's model to represent the true top of the chromosphere, then to have
a downwards ponderomotive force just below this region,
we require from equation 13 an almost zero net Alfv\'en wave energy flux
through the chromosphere, i.e. complete reflection at the
chromosphere-corona boundary to give $e=f$. In this case the wave
energy density $B_{\perp}^2/8\pi \propto\exp\left(-2z/h\right)$
and $E_{\perp}^2=V_A^2B_{\perp}^2/c^2\propto\exp\left(-z/h\right)$.
The gradient of $E_{\perp}^2$ is now directed downwards,
\begin{equation}
{\partial E_{\perp}^2\over\partial z}=-{V_A^2\over c^2}
{8\pi U_{waves}\over h}
\end{equation}
giving the ponderomotive force necessary for an inverse FIP effect.
This cancellation between these two terms is the origin of the
inverse FIP effect found above at the wave period of 408 s
corresponding to the minimum wave magnetic field.
From equation 13, we see that we require
$e-f << \left(e+f\right)\alpha ^2\exp
\left(-z/h\right)$ which for a chromosphere similar to that of the sun
($\alpha\sim 0.01$) requires cancelation between $c$
and $d$ to an accuracy of $10^{-4}$ close to $z=0$, where most
fractionation occurs in our models.

This cancelation condition requires an almost complete
reflection of waves upon first encountering the corona from below,
which can be shown analytically and numerically to practically never occur
within Hollweg's model. The
cancelation condition may only be achieved if waves are fed in from
each chromospheric footpoint with the same amplitude and phase.
This is probably more likely
in stars with lower gravity, and hence larger chromospheric scale
heights, where the degree of cancelation required is significantly lower.

A further condition required to achieve wave cancellation in the
chromosphere with waves fed in from both footpoints
is that there should be negligible wave dissipation
in the corona. \citet{hollweg84} shows that in the solar case some
coronal wave dissipation is necessary to avoid predicting unrealistically
high non thermal line broadening.
We speculate that chromospheric incident waves with
lower frequencies relative to those in the sun will need to
turbulently cascade to higher relative frequencies to dissipate if
the coronal magnetic field is higher, and that this extra decade or
two in frequency reduces the dissipation to sufficiently low levels.
Constraints on non thermal line broadening are much less stringent in stars
exhibiting the inverse FIP effect, due to their rapid rotation.

We speculate that a more realistic chromospheric density and Alfv\'en wave
profile would yield a more robust inverse FIP effect. In particular,
if the region of FIP
or inverse FIP fractionation could be identified not with the region just below
the chromosphere-corona boundary in Hollweg's model, but lower down (i.e.
$z\sim -{\rm few}\times h$ instead of $z\sim 0$) then
downwards ponderomotive forces could be more prevalent. Stars with copious
low frequency Alfv\'enic turbulence could then routinely show inverse FIP effects.
This would obviate the need for waves being fed into both loop footpoints, and
reduce the degree of cancellation required between upward and downward
propagating waves. Such turbulence is likely to derive from differential
rotation of the envelope rather than from convection, in order to provide the
necessary low frequency waves, and to match the observational constraint
that inverse FIP effects generally appear only in rapidly rotating stars.

We close this section by commenting that \citet{schwadron99} also give a
mechanism for inverse FIP effects. The combination of upwards force due to the
partial pressure of neutrals and downwards force due to the partial pressure of
ion has the effect of enhancing the coronal abundance of high FIP ions since
they ionize higher up than the low FIP elements. However quantitative
considerations indicate that such fractionation only occurs for flow speeds
in the coronal section of the loop much less than 1 km s$^{-1}$ (in section 3.1
above we discussed flow speeds through the chromosphere). At a more
realistic flow speed of 1-10 km s$^{-1}$, all fractionation disappears.

\section{Conclusions}
In this paper we have attempted to demonstrate that the ponderomotive
force arising as Alfv\'en waves propagate through the chromosphere of
the Sun or stars can give rise to the by now well documented coronal
element abundance anomalies. The upward ponderomotive force on ions that
produces the FIP effect turns out to be relatively robust for solar
parameters, and most likely diminishes as one moves either up or down in
stellar activity from the Sun. The absolute magnitude of the fractionation
produced depends sensitively on the chromospheric wave energy density.
There is nothing ``magic'' about the solar FIP fractionation of a factor
of 3-4; in the presence of higher or lower wave energy densities
higher or lower FIP fractionations will be seen. As well as providing
a successful quantitative account of the solar FIP effect and its
variations, an inverse FIP effect may also come out of the same model.
The main simplifying assumptions we have employed in this work are that the
magnetic field is constant with height in the chromospheric loop footpoints,
the Alfv\'en waves propagate only along the magnetic field (but in both
directions), and that the vertical structure of the chromosphere can be
described by the models of \citet{vernazza81} with no fractionation in the absence
of ponderomotive forces. Using observed coronal spectra
we have calculated the various element ionization fractions in the chromosphere,
and using equation 14 we have evaluated the ponderomotive acceleration and
resulting FIP fractionation. Once fully understood, coronal abundance anomalies
may offer a unique diagnostic of Alfv\'en wave propagation between the
solar/stellar chromosphere and corona.

\acknowledgements
This work has been supported by NASA Grant NAG5-9105,
by NASA Contract S13783G, and by the NRL/ONR Solar Magnetism
and the Earth's Environment 6.1 Research Option. I am indebted as ever
to my colleagues Jeremy Drake and Jim Klimchuk for encouragement,
advice and useful references.

\clearpage

\clearpage
\begin{figure}
\plotone{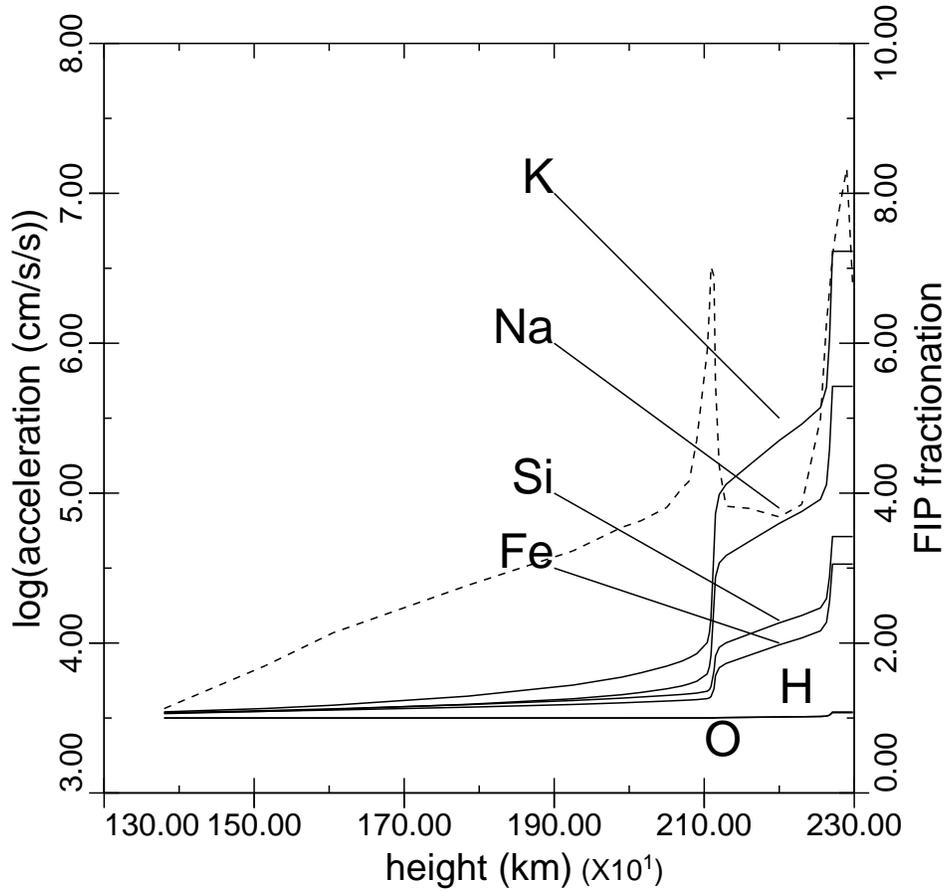}
\figcaption[f2.eps]{Plots of FIP fractionation against height (solid lines)
for O, H, Fe, Si, Na, and K, to be read on the right hand y-axis. The dashed
line gives the ponderomotive acceleration, read on the left hand y-axis. Low
FIP enhancements where this acceleration is strong can clearly be seen. The
wave energy density is 0.04 ergs cm$^{-3}$, corresponding to the quiet sun
``reference'' model.
\label{fig2}}
\end{figure}
\clearpage
\begin{figure}
\epsscale{0.5}
\plotone{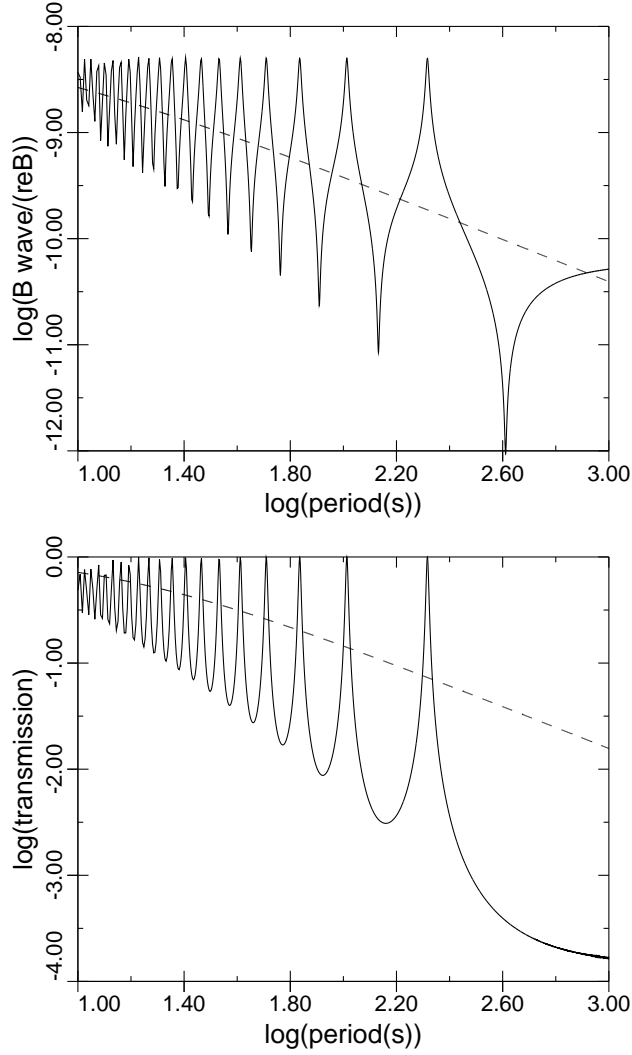}
\figcaption[f3.eps]{Plots of wave magnetic field at corona-chromosphere
boundary
(upper panel) and the Alfv\'en wave transmission into the corona
from the chromosphere (lower panel) against wave period for a model
solar chromosphere and corona (gravitational scale height $2\times 10^7$
cm, coronal Alfv\'en speed $10^8$ cm s$^{-1}$, and coronal loop length
$10^{10}$ cm). Dashed curves show corresponding properties for a coronal
hole (with loop length $\rightarrow\infty $). Strongest wave magnetic
field and hence ponderomotive force is associated with transmission
maxima. The absence of these maxima for the coronal hole leads to
relatively weak ponderomotive force and hence essentially no
FIP fractionation.}
\end{figure}
\clearpage
\begin{figure}
\plotone{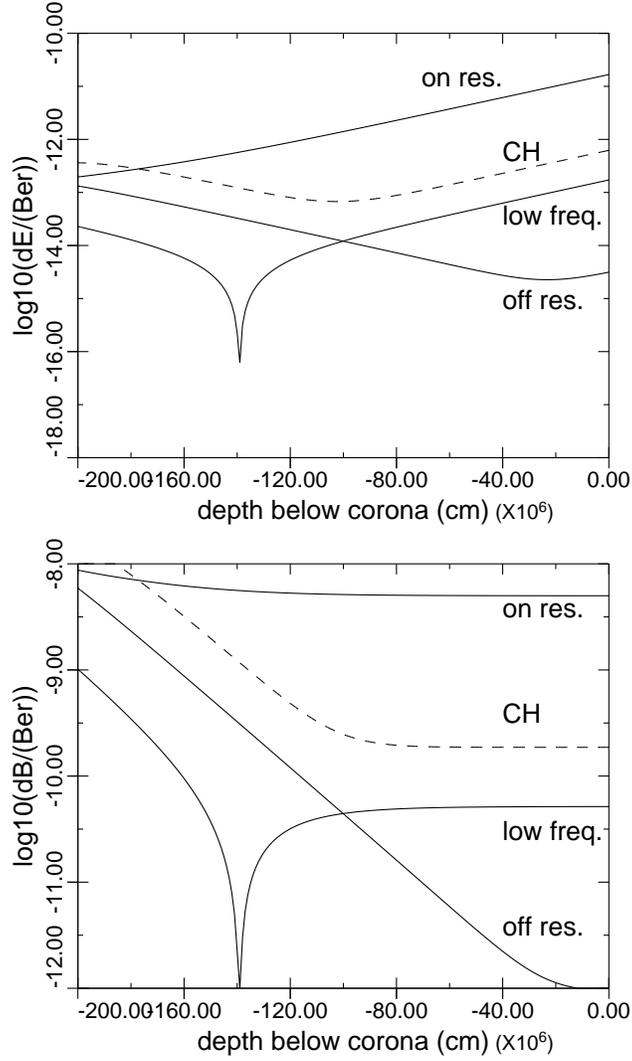}
\figcaption[f4.eps]{Plots of Alfv\'en wave electric field (upper panel)
and magnetic field (lower panel), against distance below the chromosphere
corona boundary for various wave periods; ``on res.''
corresponds to the transmission maximum at 207 s period, ``off res.''
to the transmission minimum at 408 s period, and ``low freq.'' to a
wave period of 1000 s. The ponderomotive force on ions is directed
along the gradient of the wave electric field. The ``on res.'' wave
electric field increases with height throughout the chromosphere,
giving an upwards ponderomotive force at all heights. Other wave
periods give a combination of downwards ponderomotive force at low
chromospheric heights and an upwards ponderomotive force higher up.}
\end{figure}
\clearpage

\begin{deluxetable}{cccccccc}
\tabletypesize{\scriptsize}
\tablecaption{Elastic Scattering Cross Sections with H}
\tablewidth{0pt}
\tablehead{
\colhead{element}& He& C& N& O& Ne& S& Ar}

\startdata
$\sigma _{sn}$ (10$^{-15}$ cm$^2$)& 0.89& 2.84& 2.17& 2.35& 1.05& 3.27& 1.45\\

\enddata



\end{deluxetable}

\begin{deluxetable}{cccccccccc}
\tabletypesize{\scriptsize}
\tablecaption{FIP Fractionations for VALC}
\tablewidth{0pt}
\tablehead{
\colhead{wave energy density (ergs cm$^{-3}$)}&
0.01& 0.02& 0.028& 0.04& 0.057& 0.08& 0.113& obs.\tablenotemark{a}\\
\colhead{element and FIP (eV)} &  &  & & & &  &  &\\ }

\startdata

H  (13.6)& 1.02& 1.03& 1.05& {\bf 1.07}& 1.10& 1.15& 1.21& \\
He (24.6)& 1.00& 1.00& 1.01& {\bf 1.01}& 1.01& 1.02& 1.02& \\
C  (11.3)& 1.04& 1.07& 1.10& {\bf 1.15}& 1.22& 1.33& 1.50& \\
N  (14.5)& 1.04& 1.08& 1.11& {\bf 1.16}& 1.24& 1.35& 1.54& \\
O  (13.6)& 1.02& 1.04& 1.06& {\bf 1.08}& 1.11& 1.17& 1.24& \\
Ne (21.6)& 1.02& 1.05& 1.07& {\bf 1.10}& 1.14& 1.20& 1.31& \\
Na (5.1)& 1.53& 2.33& 3.31& {\bf 5.42}& 10.9& 29.4& 119& 4-8\\
Mg (7.6)& 1.36& 1.85& 2.39& {\bf 3.42}& 5.70& 11.7& 32.5& 4\\
Al (6.0)& 1.33& 1.76& 2.23& {\bf 3.11}& 4.98& 9.69& 24.8& 4-8\\
Si (8.2)& 1.36& 1.85& 2.39& {\bf 3.42}& 5.69& 11.7& 32.4& 4\\
S  (10.4)& 1.18& 1.39& 1.59& {\bf 1.93}& 2.54& 3.73& 6.43& \\
Ar (15.8)& 1.06& 1.12& 1.17& {\bf 1.25}& 1.36& 1.55& 1.86& \\
K  (4.3)& 1.64& 2.69& 4.05& {\bf 7.22}& 16.4& 52.2& 269.& 11\\
Ca (6.1)& 1.31& 1.71& 2.13& {\bf 2.91}& 4.53& 8.46& 20.5& 4-8\\
Fe (7.9)& 1.32& 1.75& 2.20& {\bf 3.05}& 4.85& 9.32& 23.5& 4\\
Ni (7.6)& 1.27& 1.61& 1.97& {\bf 2.60}& 3.87& 6.77& 15.0& 4\\
Kr (14.0)& 1.04& 1.08& 1.11& {\bf 1.16}& 1.23& 1.34& 1.52& \\
Rb (4.2)& 1.71& 2.95& 4.62& {\bf 8.71}& 21.3& 75.8& 455.& \\
W  (8.0)& 1.72& 2.99& 4.71& {\bf 8.95}& 22.2& 80.2& 493.& \\

\enddata

\tablenotetext{a}{Estimates in the final column are observed FIP
fractionations quoted in \citet{feldman00} and \citet{feldman03}.}


\end{deluxetable}
\end{document}